\documentclass[conference]{IEEEtran}
\IEEEoverridecommandlockouts

\usepackage[utf8]{inputenc}
\usepackage{amsmath,amssymb,amsthm}
\usepackage{accents,latexsym,cancel}
\usepackage{cite}

\usepackage[dvipsnames]{xcolor}
\definecolor{myPink}{RGB}{255,105,183}

\usepackage[T1]{fontenc}
\usepackage{graphics} 
\usepackage{epsfig} 
\usepackage[mathscr]{euscript}
\usepackage{algorithm}
\usepackage[noend]{algpseudocode}
\usepackage{bbm}
\makeatletter
\def\BState{\State\hskip-\ALG@thistlm}
\makeatother

\usepackage{tikz}
\usetikzlibrary{arrows,shapes,chains,matrix,positioning,scopes,patterns,calc,
decorations.markings,
decorations.pathmorphing,
}

\usepackage{pgfplots}
\pgfplotsset{compat=1.3}
\usepgflibrary{shapes}

\renewcommand{\epsilon}{\varepsilon}

\newcommand{\RNum}[1]{\uppercase\expandafter{\romannumeral #1\relax}}

\newcommand{\av}{\ensuremath{\mathbf{a}}}
\newcommand{\bv}{\ensuremath{\underline{b}}}

\newcommand{\mv}{\ensuremath{\mathbf{m}}}

\newcommand{\rv}{\ensuremath{\mathbf{r}}}
\newcommand{\sv}{\ensuremath{\mathbf{s}}}
\newcommand{\vv}{\ensuremath{\mathbf{v}}}

\newcommand{\wv}{\ensuremath{\mathbf{w}}}
\newcommand{\xv}{\ensuremath{\mathbf{x}}}
\newcommand{\yv}{\ensuremath{\mathbf{y}}}
\newcommand{\zv}{\ensuremath{\mathbf{z}}}

\newcommand{\etav}{\ensuremath{\boldsymbol{\eta}}}

\newcommand{\Am}{\ensuremath{\mathbf{A}}}

\newcommand{\Xv}{\vec{{X}}}

\def \Mc{M_{\mathrm{c}}}

\def \Nc{N_{\mathrm{c}}}

\def \Pc{P_{\mathrm{c}}}

\def \Bc{B_{\mathrm{c}}}

\def\Pr{\mathrm{Pr}}

\DeclareMathAlphabet{\mcl}{OMS}{cmsy}{m}{n}

\newcommand{\wh}{\widehat}

\newlength\tikzwidth
\newlength\tikzheight

\definecolor{mycolor1}{rgb}{0.63529,0.07843,0.18431}%
\definecolor{mycolor2}{rgb}{0.00000,0.44706,0.74118}%
\definecolor{mycolor3}{rgb}{0.00000,0.49804,0.00000}%
\definecolor{mycolor4}{rgb}{0.87059,0.49020,0.00000}%
\definecolor{mycolor5}{rgb}{0.00000,0.44700,0.74100}%
\definecolor{mycolor6}{rgb}{0.74902,0.00000,0.74902}%

\usepackage{mathcomSTEv4}

\title{Stochastic Binning and Coded Demixing for Unsourced Random Access}
\author{\dag Jamison R. Ebert, \dag Vamsi K. Amalladinne, \S Stefano Rini, \dag Jean-Francois Chamberland, \dag Krishna R. Narayanan\\
\dag Department of Electrical and Computer Engineering, Texas A\&M University\\
\S Department of Electrical and Computer Engineering, 
National Yang Ming Chiao Tung University (NYCU)
\thanks{
This material is based upon work supported, in part, by the National Science Foundation (NSF) under Grant CCF-1619085 and by Qualcomm Technologies, Inc., through their University Relations Program.}
}

\begin{document}

\maketitle
\begin{abstract}  
Unsourced random access is a novel communication paradigm designed for handling a large number of uncoordinated users that sporadically transmit very short messages.
Under this model, coded compressed sensing (CCS) has emerged as a low-complexity scheme that exhibits good error performance. 
Yet, one of the challenges faced by CCS pertains to disentangling a large number of codewords present on a single factor graph.
To mitigate this issue, this article introduces a modified CCS scheme whereby active devices stochastically partition themselves into groups that utilize separate sampling matrices with low cross-coherence for message transmission. 
At the receiver, ideas from the field of compressed demixing are employed for support recovery, and separate factor graphs are created for message disambiguation in each cluster.
This reduces the number of active users on a factor graph, which improves performance significantly in typical scenarios.
Indeed, coded demixing reduces the probability of error as the number of groups increases, up to a point.
Findings are supported with numerical simulations. 
\end{abstract}


\section{Introduction}
\label{section:Introduction}
The Internet of Things (IoT) revolution has created a new and rapidly growing class of devices in modern communication networks: uncoordinated wireless nodes that sporadically transmit very short messages.
These devices collectively pose an important challenge as their sporadic, bursty transmissions become very costly under traditional enrollment-estimation-scheduling procedures.
As the number of such devices is envisioned to grow drastically over the next several years, there is a real and pressing need for novel communication schemes that can  handle the novel traffic pattern they generate.
In the literature, this is generally known as massive machine type communications (mMTC).
One increasingly popular model for multiple access in mMTC is the unsourced random access (URA) paradigm proposed by Polyanskiy~\cite{polyanskiy2017perspective}.
In the URA framework, currently active devices
use a common codebook at regularly scheduled slots, as facilitated by a beacon.
Because all devices share the same codebook, the signal sent by an active device depends only on its payload, not its identity.
Devices may identify themselves by inserting an identifier within their payload
which is handled by the link layer.
%
The receiver is thus tasked with producing an unordered list of all transmitted messages.
The predominant performance criterion used to evaluate an URA scheme is the per-user probability of error (PUPE), i.e., the probability that a transmitted message is not present in the recovered list.

Several pragmatic schemes have been shown to exhibit low PUPE over SNRs of interest \cite{vem2019user,amalladinne2020coded,calderbank2019chirrup,fengler2019sparcs-isit,amalladinne2020AMP,andreev2020polar,pradhan2020polar}, and these solutions vary in performance, computational efficiency and sampling complexity.
A noteworthy URA scheme is coded compressed sensing (CCS)~\cite{amalladinne2020coded}.
Therein, Amalladinne et al.\ treat joint decoding as an instance of the noisy support recovery problem, albeit one whose dimensionality precludes the straightforward use of standard compressed sensing (CS) solvers.
The CCS approach breaks the CS problem into several sub-problems of tractable dimensions.
Each section is encoded using traditional CS techniques and can be decoded via established algorithms such as approximate message passing (AMP).
An outer code is then leveraged to stitch message portions together.
CCS has been shown to perform well, and it has been extended to accommodate more general designs \cite{calderbank2019chirrup,fengler2019sparcs-isit,amalladinne2020AMP}.

Recently, the CCS scheme was adapted to serve multiple classes of devices~\cite{amalladinne2020multiclass}: under this alternate formulation, devices with different transmit powers or message lengths can be processed simultaneously over a same channel.
To achieve concurrent transmission, each class of users employs its own dictionary for CS encoding, and compressed demixing techniques are subsequently used to recover the messages from each class \cite{bhaskar2013atomic,mccoy2014convexity,mccoy2014sharp}.
This article improves the state-of-the-art CCS formulation in \cite{amalladinne2020AMP} by incorporating ideas from multi-class CCS \cite{amalladinne2020multiclass}, where sampling matrices with low cross-coherence are utilized to separate classes.
Specifically, we demonstrate that significant performance improvements can be obtained by having active devices stochastically partition themselves into groups before encoding their messages.
Every group then uses its own dictionary for inner encoding, paralleling the development of coded demixing.
One distinction of this approach compared to multi-class CCS arises from the fact that group sizes are unknown beforehand, and hence the number of devices within a group must be estimated.
Performance benefits are significant, as evinced by numerical simulations.

\noindent
{\bf Notation:} Matrices are denoted with bold capital letters; column vectors, with lowercase ones.
The set of integers $\{1,\ldots,N\}$ is denoted by $[N]$.
%
We write the $\ell$-norm of $\av$ as $\| \av \|_\ell$. 
The concatenation of two vectors is $\vv\uv \triangleq [\vv^\intercal \ \uv^\intercal]^\intercal$.

\section{System Model}
\label{section:SystemModel}

We immediately turn to the problem formulation.
In the URA model we examine, the signal received at the destination over $n$ channel uses is obtained as
\begin{equation} \label{equation:ChannelModel}
\textstyle 
\yv = 
\sum_{i \in [K]} 
\xv_i + \zv,
\end{equation}
where $\xv_i \in \mathbb{R}^n$ is the signal sent by active device~$i \in [K]$ with $\|\xv_i\|_2 \leq nP~\forall~i \in [K]$.
Parameter $K$ denotes the number of active devices, and $i$ is an arbitrary, yet fixed labeling of these devices.
Additive noise $\zv$ in \eqref{equation:ChannelModel} is Gaussian with $n$ independent $\mathcal{N}(0,1)$ components.
Following established URA literature, the transmitted signal $\xv_i$ is a function of the information message at device~$i$, but it is impervious to the identity of the device itself.
We mention briefly that the challenge of designing a communication scheme for this mathematical model is equivalent to that of formulating a CS scheme, albeit one whose dimension (on the order of $2^{128}$ in our case) precludes the application of standard CS solvers.
Thus, the task at hand is to design a pragmatic communication scheme that recovers the transmitted messages with high probability.

Overall performance for a URA implementation is typically assessed using the PUPE~\cite{polyanskiy2017perspective}.
That is, the access point is asked to deliver an unordered list $\widehat{W}(\yv)$ of (at most) $K$ messages based on observed signal $\yv$.
This evaluation criterion is expressed mathematically as
\begin{equation} \label{equation:PUPE}
P_{\mathrm{e}} = \f {1}{K}
\textstyle  
\sum_{i \in [K]}
\Pr \Big( \wv_i \notin \widehat{W}(\yv) \Big),
\end{equation}
where $\wv_i$ is the payload of active device~$i$.
We note that, in most of the literature, $K$ is given as side information; we adhere to this convention.
In practice, it may need to be estimated as a preliminary step of the frame establishment process.
The number of channel uses available is roughly $n \approx 30,000$. 

\section{Proposed Approach}
\label{section:ProposedApproach}

The conceptual starting point for our discussion is the CCS scheme of Amalladinne et al.~\cite{amalladinne2020AMP,amalladinne2020unsourced}.
This URA approach combines an LDPC outer code and a CS-style inner code reminiscent of a sparse regression code (SPARC)~\cite{rush2017capacity,CIT-092}.
More formally, an information message $\wv$ is divided into sections of length $v$, each of which may be viewed as an element of the Galois field $\mathbb{F}_2^v$. 
The message is then encoded using a non-binary LDPC code, yielding codeword $\vv \in \{0, 1\}^{v \times L}$, which is expressed as the concatenation:
\begin{equation} \label{equation:ConcatenatedFragments}
\vv = \vv(1) \vv(2) \cdots \vv(L) .
\end{equation}
The next step in the encoding process is to map each section into its index representation, $\vv (\ell) \mapsto \mv(\ell)$. 
Here, $\mv (\ell)$ is a length-$2^v$ vector with zeros everywhere except a location $\left[\vv(\ell)\right]_2$, where $\left[\vv(\ell)\right]_2$ is an integer expressed in binary form. 
The concatenated vector $\mv = \mv(1) \cdots \mv(L)$ is a SPARC-like codeword with $L$ non-zero locations.
The transmitted signal then assumes the form $\xv = d \Am \mv$ for a prescribed sensing matrix $\Am$, where $d$ is a scalar value capturing signal amplitude.

\begin{figure}
\centering
\begin{tikzpicture}[
  font=\footnotesize, >=stealth', line width = 0.75pt,
  fragment/.style={rectangle, minimum height=6mm, minimum width=35mm, draw=black, fill=gray!10, rounded corners},
  block/.style={rectangle, minimum height=6mm, minimum width=15mm, draw=black, fill=gray!40, rounded corners}
]

\node[fragment] (fragment) at (0,2.25) {Information Message};

\foreach \j in {1,2} {
    \node[block] (block\j) at (1.5*\j-3.75,1) {$\vv(\j)$};
}
\node at (0.75,1) {$\cdots$};
\node[block] (block4) at (2.25,1) {$\vv(L)$};

\draw (fragment.south west) to (block1.north west);
\draw (fragment.south east) to (block4.north east);
\node (encoding) at (0,1.625) {Outer Encoding};

\node[rotate=45] (index) at (0.375,-0.2) {Indexing};

\draw[->] (block1.south) to [out=-45,in=135] (-1.875,0.25);
\draw[->] (block2.south) to [out=-90,in=0] (-1.25,-0.25);
\node (encoding) at (-2.25,-0.5625) {$\xv = d\Am \left( \; \begin{bmatrix} \mv(1) \\ \mv(2) \\ \vdots \\ \mv(L) \end{bmatrix} \; \right)$};
\draw[->] (block4.south) to [out=-105,in=0] (-1.25,-1.25);

\end{tikzpicture}
\caption{This illustration encapsulates encoding for CCS-AMP.
Redundancy is added to message $\wv$ in the form of an outer code.
Codeword $\vv$ is then split into blocks, which are indexed individually.
The resulting SPARC-like vector is processed using sensing matrix $\Am$.}
\label{figure:CodedCompressedSensing}
\end{figure}
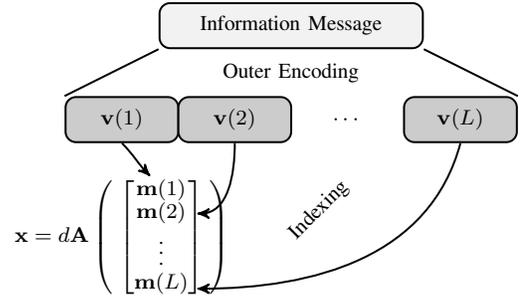
This particular CCS construction is amenable to iterative decoding via the AMP algorithm~\cite{fengler2019sparcs-isit}.
Furthermore, the AMP inner decoder can take advantage of the LDPC outer code structure within the denoising step.
This improves performance while maintaining a similar computational complexity~\cite{amalladinne2020AMP,amalladinne2020unsourced}.
It is pertinent to note that the outer code plays two crucial roles: it boosts convergence for the AMP algorithm, and it enables codeword disambiguation once AMP has converged.
An important distinction in the former use of an LDPC code is that several codewords reside on the same bipartite graph.
This forces the LDPC alphabet size to be large, as to avoid collisions and to permit belief propagation based on marginal distributions~\cite{amalladinne2020unsourced}.
This situation also demands other characteristics such as the presence of check nodes with small degrees, and the application of efficient techniques such as the Fast Walsh–Hadamard transform to carry out message passing.

One challenge associated with such implementations stems from the fact that several codewords reside on the same factor graph within the outer code.
This muddling of codewords renders the application of belief propagation more complex, and it impacts performance negatively.
A naive explanation for this phenomenon is that local factors cannot distinguish symbols associated with separate codewords.
This results in a type of local mixing that clouds the decoding process.
To circumvent this impediment, we explore stochastic binning as a means to partition active devices in different groups.

\subsection{Coded Demixing}

In our proposed approach, an active device utilizes the first $w_0$ bits of its payload to select a sensing matrix, which is conceptually attached to a bin.
We note that the bin assignment process cannot be deterministic as the set of active devices is unknown within the URA framework. 
As we will see later, we dedicate a few channel uses for bin occupancy estimation; but for ease of presentation, we overlook these details for the time being.
The device encodes the remainder of its message using the CCS paradigm described above.
This binning process reduces the effective number of codewords residing on a same outer factor graph, a highly desirable outcome.
This comes at a small cost in computational complexity.
It also changes the operating point of the CS system in terms of sparsity and undersampling ratio.
Since AMP can be sensitive to these attributes, design parameters must be picked carefully.
We expound on this process below.

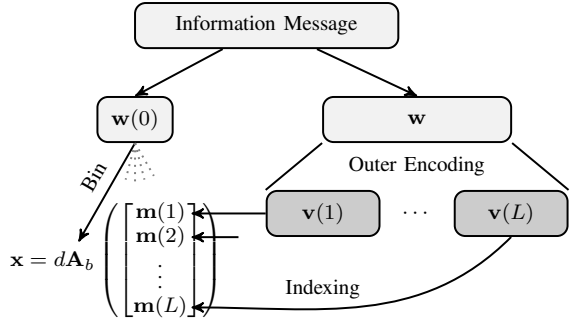
\begin{figure}
\centering
\begin{tikzpicture}[
  font=\footnotesize, >=stealth', line width = 0.75pt,
  fragment/.style={rectangle, minimum height=6mm, minimum width=25mm, draw=black, fill=gray!10, rounded corners},
  block/.style={rectangle, minimum height=6mm, minimum width=15mm, draw=black, fill=gray!40, rounded corners}
]

\node[fragment,minimum width=35mm] (message) at (-1,3.5) {Information Message};

\node[fragment,minimum width=10mm] (fragment0) at (-2.75,2.25) {$\wv(0)$};
\node[fragment] (fragment) at (1,2.25) {$\wv$};
\draw[->] (message) to (fragment0.north);
\draw[->] (message) to (fragment.north);

\node (encoding) at (1,1.625) {Outer Encoding};
\node[block] (block1) at (-0.25,1) {$\vv(1)$};
\node at (1,1) {$\cdots$};
\node[block] (block4) at (2.25,1) {$\vv(L)$};

\draw (fragment.south west) to (block1.north west);
\draw (fragment.south east) to (block4.north east);

\node[rotate=60] (bin) at (-3.3,1.5) {Bin};
\node (index) at (-0.25,0) {Indexing};

\draw[->] (fragment0.south) to (-3.5,0.625);
\draw[dotted, draw=gray] (fragment0.south) to (-2.875,1.45);
\draw[dotted, draw=gray] (fragment0.south) to (-2.75,1.4);
\draw[dotted, draw=gray] (fragment0.south) to (-2.625,1.45);
\draw[dotted, draw=gray] (fragment0.south) to (-2.5,1.5);

\draw[->] (block1.west) to (-2,1);
\draw[->] (-1.375,0.685) to (-2,0.685);
\node (encoding1) at (-3,0.375) {$\xv = d\Am_b \left( \begin{bmatrix} \mv(1) \\ \mv(2) \\ \vdots \\ \mv(L) \end{bmatrix} \right)$};
\draw[->] (block4.south) to [out=-135,in=0] (-2,-0.25);

\end{tikzpicture}
\caption{In coded demixing, the first few bits $\wv(0)$ of a message are employed to select a bin and a sensing matrix $\Am_b$.
The remaining information bits $\wv$ are encoded using standard CCS.
This illustration encapsulates encoding for CCS-AMP.
This produces a statistical partitioning across bin akin to demixing.}
\label{figure:CodedDemixing}
\end{figure}
Let $\{ \Am_b \}$ be a collection of $n \times m$ matrices over the real numbers, with $b \in [B]$ and $B = 2^{w_0}$.
To send information, a device employs the first $w_0$ bits of its payload to select bin index~$b$, where $b$ is the integer value of binary sequence $\wv(0)$ (plus one).
The device takes the remaining bits to create a codeword under the CCS-style encoding described at the beginning of Section~\ref{section:ProposedApproach}.
That is, transmitted signal $\xv_i$ is obtained via the product $\xv_i = d \Am_b \mv_i$ over the field of real numbers, where $\mv_i$ is the CCS index representation of the residual message, omitting the bin selection portion $\wv(0)$.
With all the active devices utilizing a similar approach, the process yields a received vector $\yv$ of the form
\begin{equation} \label{equation:BCCS}
\begin{split}
\yv &= \textstyle \sum_{b \in [B]} \sum_{i \in \mathcal{G}_b} d \Am_b \mv_i + \zv 
= \textstyle \sum_{b \in [B]} d \Am_b \sv_b + \zv
\end{split}
\end{equation}
where $\sv_b = \sum_{i \in \mathcal{G}_b} \mv_i$ and $\mathcal{G}_b$ is a shorthand notation for the collection of active devices that have selected bin~$b$; necessarily, $\bigcup_{b} \mathcal{G}_b = [K]$.
The demixing interpretation of \eqref{equation:BCCS} stems from the task of recovering $\sv_1, \ldots, \sv_B$ from $\yv$.

\subsection{Inner Code and AMP Equations}

If we create a dictionary of possible signals across bins as
\begin{equation} \label{equation:Dictionary}
\Am = \begin{bmatrix} \Am_1 & \Am_2 & \cdots & \Am_B \end{bmatrix},
\end{equation}
and we concatenate state vectors $\sv_1, \ldots, \sv_B$ into tall vector $\sv$, then we can again express the received signal as
\begin{equation} \label{equation:LinearSystem}
\yv = d \Am \sv + \zv .
\end{equation}
This equation conforms to the standard CS form, although system properties differ.
The undersampling fraction is $\delta = \frac{n}{B \cdot L \cdot 2^v}$, and the measure of sparsity is $\rho = \frac{K \cdot L}{n}$.
The number of demixing bins, the alphabet size of the outer code, and the rate of the outer code all play a role in establishing the sparsity-undersampling operating point of the resulting system.

Equation~\eqref{equation:LinearSystem} admits a two-step AMP iteration process to recover sparse vector $\sv$.
For $t=0, 1, \ldots$, this algorithm alternates between the following two equations:
\eas{
&\zv^{(t)} = \yv - d \Am \sv^{(t)} + \frac{\zv^{(t-1)}}{n}
\operatorname{div} d \etav^{(t-1)} \big( \rv^{(t-1)} \big)
\label{equation:AMP} \\
&\sv^{(t+1)} = \etav^{(t)} \big( \rv^{(t)} \big),
\label{equation:denoiser}
}
where the effective observation is $\rv^{(t)} = d \sv^{(t)} + \Am^{\intercal} \zv^{(t)}$ and $\Am^{\intercal}$ denotes the transpose of matrix $\Am$.
The algorithm is initialized with vectors $\zv^{(-1)} = \mathbf{0}$ and $\sv^{(0)} = \mathbf{0}$.
The first equation can be interpreted as the computation of the residual signal enhanced by an Onsager correction \cite{bayati2011dynamics,donoho2013information}.
The denoiser in \eqref{equation:denoiser} produces an estimate for $\sv$ at iteration $t+1$, and it seeks to take advantage of known signal attributes.
This structure is very much in the spirit of AMP for sparse regression codes~\cite{rush2017capacity,greig2017techniques} and prior instances of CCS-AMP~\cite{fengler2019sparcs-isit,amalladinne2020AMP,amalladinne2020unsourced}.

\subsubsection{Occupancy Estimation}

The denoiser we adopt is similar to the CCS-AMP formulation~\cite{amalladinne2020AMP,amalladinne2020unsourced}.
In view of \eqref{equation:Dictionary} and \eqref{equation:LinearSystem}, we can think of the effective observation $\rv$ as a concatenation of $B$ vectors of the form
$\rv_b^{(t)} = \Am_b^{\intercal} \zv_b^{(t)} + d \sv_b^{(t)}$.
If the number of devices that selected bin~$b$, say $K_b$, were available as side information, then we would readily be able to apply the standard CCS-AMP denoiser to this component of the problem.
The interference from other bins would appear as additive noise, and the number of codewords on the factor graph of the outer code associated with bin~$b$ would be greatly reduced, with a high probability.
Unfortunately, we do not have access to $\{ K_b : b \in [B] \}$ a priori.
Instead, we must produce estimates $\{ \Kh_b : b \in [B] \}$.
We turn to a description of the estimation process next.  


Before transmitting their respective CCS-like messages, every device sends a length $2^{w_0}$ standard basis vector $\xv(0)$ where the single unity entry corresponds to the index $b$ of the chosen bin.
The receiver then employs a standard estimator to create bin cardinality estimates $\{ \Kh_b : b \in [B] \}$ for the $B = 2^{w_0}$ bins.
We note that, under the URA formulation, the expected energy dissipated by a device must be no greater $n P$.
The amount of energy per bit dedicated to bin occupancy must therefore be subtracted from the total energy budget.
Likewise, the number of channel uses associated with $\xv(0)$ must also be accounted for.
These idiosyncrasies should be straightforward to the reader familiar with URA and, hence, details are omitted.
Nonetheless, the simulations presented below in Sec.  \ref{sec:Design Consideration and Simulations} properly account for the energy and number of channel uses dedicated to bin occupancy estimation.


\subsubsection{Dynamic Denoiser}

Our denoising function operates separately on distinct bins.
Within a bin, it essentially takes the form of the denoiser found in~\cite{amalladinne2020AMP,amalladinne2020unsourced}, which leverages the LDPC outer code structure at every iteration.
Given effective observation $\rv_b$, the algorithm initializes the beliefs on the factor graph of the outer code using the marginal posterior mean estimate (PME) of Fengler et al.~\cite{fengler2019sparcs-isit,fengler2019sparcs}.
That is, for every individual component, we get
\begin{equation} \label{equation:OriginalPME}
\hat{s}_{b,\ell} \left( r, \tau \right)
= \frac{q e^{ - { \left( r - d \right)^2}/{2 \tau^2} }}
{(1-q) e^{ -{r^2}/{2 \tau^2} }
+ q e^{ - { \left( r - d \right)^2}/{2 \tau^2} }},
\end{equation}
where 
$q = 1-\left(1-2^{-v}\right)^{\Kh_b}$ is a constant.
Note that we have made dependencies on $\ell$, $b$, and $t$ implicit in many instances for ease of exposition.
Furthermore, \eqref{equation:OriginalPME} utilizes estimate $\Kh_b$ rather than the exact value $K_b$, which is not available.
The overall bin denoiser is the concatenation
\begin{equation}
\label{equation: bindenoiser}
\hat{\sv}_b \left( \rv_b, \tau \right)
= \hat{\sv}_{b,1} \left( \rv_b(1), \tau \right) \cdots
\hat{\sv}_{b,L} \left( \rv_b(L), \tau \right),
\end{equation}
where the $\kappa$th element of $\hat{\sv}_{b,\ell} \left( \rv_b(\ell), \tau \right)$ is obtained by computing $\hat{s}_{b,\ell} ( \cdot, \tau )$ with the $\kappa$th element of $\rv_b(\ell)$ as its argument.
The denoiser then performs one round of belief propagation on the factor graph of the outer code using standard message passing rules.
This step seeks to account for the connections between neighboring blocks in the outer code through local factors.
After clipping and rescaling, this procedure yields an estimate for the support of $\sv_b$.

The overall denoiser in \eqref{equation:denoiser} (neglecting transposition) is
\begin{equation}
\label{equation: overalldenoiser}
\etav^{(t)} \big( \rv \big) = \hat{\sv}_1 \left( \rv_1, \tau_t \right) \hat{\sv}_2 \left( \rv_2, \tau_t \right)
\cdots \hat{\sv}_B \left( \rv_B, \tau_t \right) .
\end{equation}
We emphasize that, although the denoiser acts on individual sections, it is non-separable.
Thus, this specific AMP implementation subscribes to the extended framework for non-separable functions characterized by Berthier, Montanari, and Nguyen~\cite{berthier2017state}.
It is also instructive to point out that the operations in \eqref{equation:denoiser} can be performed in parallel across bins.
Thus, the resulting complexity increase in the denoising step is linear in the number of bins created.

\subsubsection{Onsager Correction}

Under the strategy of first estimating bin occupancy and then demixing, the Onsager term in \eqref{equation:AMP} can be computed in a straightforward fashion.
Indeed, $\frac{\partial \hat{K}_{b, \ell}}{\partial \rv_b(\ell, \kappa)} = 0$ because section~$\ell$ does not enter the computation of $\hat{K}_{b, \ell}$.
Consequently, the partial derivative with respect to $r$ of the posterior mean estimator (PME) remains
\begin{equation}
\frac{\partial\hat{s}_{b, \ell}(q, r, \tau)}{\partial r}
= \frac{d}{\tau^2}\hat{s}_{b, \ell}(q, r, \tau)
\left( 1 - \hat{s}_{b, \ell}(q, r, \tau) \right),
\end{equation}
as in~\cite{amalladinne2020unsourced}.
Consequently, the divergence of $d \etav^{(t)} \left( \rv \right)$ with respect to $\rv$ is equal to
\begin{equation} \label{equation:BP-PME-OnsagerCorrection}
\operatorname{div} d \etav^{(t)} \left( \rv \right)
= \frac{d^2}{\tau_t^2} \left( \left\| \etav^{(t)} \left( \rv \right) \right\|_1 - \left\| \etav^{(t)} \left( \rv \right) \right\|_2 ^2 \right).
\end{equation}
The complexity order for the computation of the residual in \eqref{equation:AMP} is also linear in the total number of bins.
In addition, these computations can be performed in parallel before aggregating the various components in producing $\zv^{(t)}$.

We conclude this section by stating that, although $\tau$ is a deterministic quantity and can be computed through the state evolution~\cite{bayati2011dynamics,berthier2017state}, it is often approximated in practice as $\tau_t^2 \approx \left\| \zv^{(t)} \right\|_2 ^2/n$.
This is the way we approach simulations.

\subsection{Codeword Disambiguation}

Once the AMP iteration process has converged, candidate codewords must be recovered from $\hat{\sv}$.
This again takes place with the decoder operating on individual bins.
Specifically, the top $\hat{K_b} + \delta$ LDPC codeword candidates in every section are identified by running several instances of belief propagation on $\hat{\sv}_b$.
The likelihood of each LDPC codeword is computed and cataloged.
The bin identifier $\wv(0)$ is prepended to the information sequence corresponding to each LDPC codeword retained.
These possible messages are aggregated across bins and ranked according to their likelihood values, starting with the most likely one.
If the list of possible messages exceeds the prescribed limit, then it is truncated to $K$ items.
The retained messages are returned as $\widehat{W}(\yv)$.

\section{Design Consideration and Simulations}
\label{sec:Design Consideration and Simulations}
The main goal of this section is to showcase the potential benefits of coded demixing for single-class URA.
We adopt system parameters aligned with the URA literature~\cite{polyanskiy2017perspective,vem2019user,amalladinne2020coded,calderbank2019chirrup,fengler2019sparcs-isit,amalladinne2020AMP,andreev2020polar} to facilitate an easy comparison with other CCS schemes.
It is worth mentioning that these parameters are inspired by the LoRaWAN wireless specifications.
There are $K = 64$ total active devices in the system, and each active device seeks to transmit a $w = 128$~bit message over $n = 38400$ channel uses.
The energy per bit is defined as $\frac{E_b}{N_0} = \frac{n P}{2w}$, where $P$ is the total energy budget.
Each sensing matrix $\Am_b$ is obtained by randomly selecting $n$ rows of a $m \times m$ Hadamard matrix. 

Out of the available channel uses, exactly $2^{w_0}$ of them are dedicated to occupancy estimation.
This is a minute amount, yet this phase of the transmission process ensures that good estimates $\{ \Kh_b : b \in [B] \}$ are available at the access point.
The amount of power apportioned to this task represent roughly $0.2B \%$ of the total energy budget.
This leads to an equivalent observation $y_b$ for every bin distributed according to $\mathcal{N}(d_0 K_b, 1)$.
An LMMSE estimator then produces $\{ \Kh_b : b \in [B] \}$ where, for this problem, $\hat{\mathbf{K}}_{1:B} = \mathbf{C}^\intercal \yv$ with $\mathbf{C}$ being the standard LMMSE matrix.

A naive interpretation of binning would suggest that increasing the number of bins is likely to improve performance.
Yet, we turn to the notions of undersampling fraction $\delta$ and the measure of sparsity $\rho$ found in \cite{donoho2009message} to emphasize design considerations.
Overlooking the estimation phase, the sparsity level for our implementation is $\rho = \textstyle \sum_b K_b L / n = {K L}/{n}$. 
The undersampling fraction, on the other hand, can be expressed as
$\delta = {n} / {B L 2^v}$.
Augmenting the number of bins drives the undersampling fraction towards zero.
Even without binning, the CCS sparsity–undersampling operating point $(\delta, \rho)$ is typically close to $(0,0)$.
Admitting more bins further drives $\delta$ towards $0$, while $\rho$ remains fixed.
This may degrade the performance of AMP in the presence of noise and it may limit the number of bins that can be reliably utilized.
Although the alphabet size of the outer code may be reduced to increase $\delta$, this would require other parameters to increase.  
Thus, it is not clear how to select the optimal combination of system parameters. 
Additionally, AMP performance seems somewhat sensitive to the rate of the outer code.
Altogether, while the parameter space is vast, tuning parameters appears to be a delicate endeavor.

For the reasons above, the parameters we adopt are essentially the same as the ones found in \cite{amalladinne2020AMP}.
Specifically, the outer code for every bin is a triadic LDPC code with $L=16$ and rate $1/2$.
This produces an alphabet size equal to $2^v = 65536$.
The number of bins ranges from one (original CCS scheme) to $B = 8$.
We note that schemes with a larger number of bins can transmit an additional $w_0 = \log_2 B$ bits, yet we neglect this marginal benefit for the sake of exposition.
The combined effect of bypassing reparametrization and overlooking extra bits therefore produces a conservative assessment of the gains achieved through coded demixing.

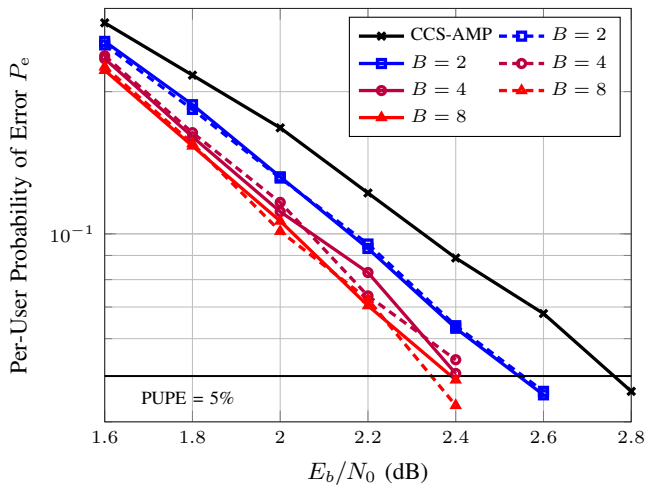
\begin{figure}[t]
\centerline{\begin{tikzpicture}

\begin{semilogyaxis}[
font=\small,
width=7cm,
height=5.5cm,
scale only axis,
every outer x axis line/.append style={white!15!black},
every x tick label/.append style={font=\scriptsize},
xmin=1.6,
xmax=2.8,
xtick={1.6, 1.8, 2.0, 2.2, 2.4, 2.6, 2.8},
xlabel={$E_{b}/N_0$ (dB)},
xmajorgrids,
xminorgrids,
every outer y axis line/.append style={white!15!black},
every y tick label/.append style={font=\scriptsize},
ymin=0.04,
ymax=0.3,
ytick={0.01, 0.1, 1.0},
ylabel={Per-User Probability of Error $P_{\mathrm{e}}$},
ymajorgrids,
yminorgrids,
legend style={anchor=north east,draw=black, fill=white, legend cell align=left,font=\scriptsize, legend columns=2}
]

\addplot [color=black,line width=1.25pt,mark=x]
  table[row sep=crcr]{
  1.6 0.27968749999999987 \\
  1.8 0.21671875000000004 \\
  2.0 0.16765624999999995 \\
  2.2 0.12203124999999998 \\
  2.4 0.08890624999999995 \\
  2.6 0.06781249999999996 \\
  2.8 0.046406089 \\
};
\addlegendentry{CCS-AMP};

\addplot [color=blue,densely dashed,line width=1.25pt,mark=square,mark size=1.5pt,mark options={solid}]
  table[row sep=crcr]{
  1.6 2.510937500000000600e-01 \\
  1.8 1.834374999999999478e-01 \\
  2.0 1.312499999999999500e-01 \\
  2.2 9.515624999999992839e-02 \\
  2.4 6.390624999999997002e-02 \\
  2.6 0.046406249999999996 \\
};
\addlegendentry{$B=2$};

\addplot [color=blue,line width=1.25pt,mark=square,mark size=1.5pt,mark options={solid}]
  table[row sep=crcr]{
  1.6 2.554687500000001887e-01 \\
  1.8 1.876562499999999689e-01 \\
  2.0 1.320312499999999889e-01 \\
  2.2 9.312499999999990230e-02 \\
  2.4 6.312499999999994504e-02 \\
  2.6 4.562499999999997113e-02 \\
};
\addlegendentry{$B=2$};

\addplot [color=purple,densely dashed,line width=1.25pt,mark=o,mark size=1.5pt,mark options={solid}]
  table[row sep=crcr]{
  1.6 2.381250000000001421e-01 \\
  1.8 0.16374999999999995 \\
  2.0 1.165624999999999439e-01 \\
  2.2 7.390624999999992339e-02 \\
  2.4 0.0541424358 \\
};
\addlegendentry{$B=4$};

\addplot [color=purple,line width=1.25pt,mark=o,mark size=1.5pt,mark options={solid}]
  table[row sep=crcr]{
  1.6 2.348437500000000455e-01 \\
  1.8 1.604687499999999933e-01 \\
  2.0 1.114062499999999428e-01 \\
  2.2 8.281249999999992784e-02 \\
  2.4 0.050624999999999976 \\
};
\addlegendentry{$B=4$};

\addplot [color=red,densely dashed,line width=1.25pt,mark=triangle,mark size=1.5pt,mark options={solid}]
  table[row sep=crcr]{
  1.6 2.256250000000000755e-01 \\
  1.8 1.553125000000000477e-01 \\
  2.0 1.012499999999999373e-01 \\
  2.2 0.07265624386 \\
  2.4 0.04332035 \\
};
\addlegendentry{$B=8$};

\addplot [color=red,line width=1.25pt,mark=triangle,mark size=1.5pt,mark options={solid}]
  table[row sep=crcr]{
  1.6 2.218750000000000444e-01 \\
  1.8 1.534375000000000322e-01 \\
  2.0 1.064062499999998967e-01 \\
  2.2 7.046874999999995504e-02 \\
  2.4 4.906249999999998113e-02 \\
};
\addlegendentry{$B=8$};

\draw[black, line width=0.75pt] (axis cs:1.6,0.05) to (axis cs:2.8,0.05);
\node at (axis cs:1.79,0.045) {\scriptsize PUPE = 5\%};

\end{semilogyaxis}
\end{tikzpicture}}
  \caption{This plot illustrates the potential benefits of coded demixing.
  Performance improves with the number of demixing bins.
  Dashed lines are genie-aided lower bounds, and solid lines represent actual performance.}
  \label{fig:pupe_by_binsize}
\end{figure}
Results are illustrated in Fig.~\ref{fig:pupe_by_binsize}, with $E_{b}/N_0$ acting as the free variable.
Two curves are present for every bin count: the genie-aided performance is a lower bound where the access point is given the exact occupancy in every bin (dashed lines); and the actual coded demixing reports the PUPE for systems where the number of codewords per bin must be inferred as part of the decoding process (solid lines).
AMP is run for $10$ iterations per simulation, with only one round of belief propagation on the outer graph per denoising step.
Each point is averaged over $100$ trials for statistical accuracy. 

The performance benefits derived from coded demixing are significant, with a $0.4$~dB reduction in $E_b/N_{0}$ when going from one bin to eight bins at a PUPE of five percent.
It appears that additional gain may be realized by pursuing bin sizes greater than $B = 8$ bins, yet this would requires additional computations.
The limit of how many bins may be employed before AMP fails to converge remains an open research question.  
It is also not known how much the performance of coded demixing may be improved through the use of alternate sensing matrices with better cross-coherence properties. 
Finally, another open problem is finding the optimal signaling and inference scheme for estimating $\{ \Kh_b : b \in [B] \}$ at the access point. It may be possible to perform these tasks jointly.

\section{Discussion}

Coded demixing offers an excellent trade-off between performance and complexity.
Yet, it is important to note that this approach does not outperform certain spread URA schemes in terms of PUPE, e.g., \cite{andreev2020polar,pradhan2020polar}.
Nevertheless, it represents the state-of-the-art in CCS, a family of URA schemes with lower complexity than those cited above.
Also, there is an interesting connection between spread URA schemes and coded demixing.
In spread URA, users pick spreading sequences as bins to limit interference while decoding; in coded demixing, users select a sensing matrix and a factor graph over which to encode and transmit their messages.  
Both schemes achieve significant performance improvements by lowering the number of users per bin via the stochastic grouping of active users.

\bibliographystyle{IEEEbib}
\bibliography{IEEEabrv,spawc2021}

\end{document}